\documentclass[twocolumn,showpacs,preprintnumbers,amsmath,amssymb,eqsecnum]{revtex4}

\usepackage{graphicx}
\usepackage{dcolumn}
\usepackage{bm}
\usepackage{amsthm,amscd,latexsym,mathrsfs}

\begin{document}
\preprint{APS/123-QED}

\title{On Quantum Field Theories in Operator and Functional Integral Formalisms}

\author{Aba Teleki}%
 \email{ateleki@ukf.sk}
 \affiliation{Department of Physics, Faculty of Natural
 Sciences,\\
 Constantine the Philosopher University, %
 SK-949 74 Nitra, Slovakia}
\author{Milan Noga}%
 \email{Milan.Noga@fmph.uniba.sk}
\affiliation{Department of Theoretical Physics,
Faculty of
Mathematics,
Physics and Computer Sciences,\\
Comenius University, %
SK-842 48 Bratislava, Slovakia}%



\date{\today}

\begin{abstract}
Relations and isomorphisms between quantum field theories in
operator and functional integral formalisms are analyzed from the
viewpoint of inequivalent representations of commutator or
anticommutator rings of field operators. A functional integral in
quantum field theory cannot be regarded as a Newton-Lebesgue
integral but rather as a formal object to which one associates
distinct numerical values for different processes of its
integration. By choosing an appropriate method for the integration
of a given functional integral, one can select a single
representation out of infinitely many inequivalent representations
for an operator whose trace is expressed by the corresponding
functional integral. These properties are demonstrated with two
exactly solvable examples.
\end{abstract}

\maketitle

\section{\label{s:introduction}Introduction}
In quantum field theory based on the functional integral formalism
\cite{Berezin,FaddeevSlavnov,Popov,Vasiliev,ChaichianDemichev}, the
partition function $\mathscr{Z}$ and quantum expectation values
$\langle {A}\rangle$ of physical observables $A$ are given by the
formal functional integrals
\begin{subequations}\label{e:fundamental}
\begin{equation}\label{e:functional Z}
    \mathscr{Z}=\int e^{-S(a_{\vphantom{k}}^{*}, a)}\mathscr{D}(a_{\vphantom{k}}^{*}, a)
\end{equation}
and
\begin{equation}\label{e:<A>}
    \langle A\rangle=\frac{1}{\mathscr{Z}}\int
    A(a_{\vphantom{k}}^{*}, a)e^{-S(a_{\vphantom{k}}^{*}, a)}\mathscr{D}(a_{\vphantom{k}}^{*}, a),
\end{equation}
\end{subequations}
respectively, where $S(a_{\vphantom{k}}^{*}, a)$ is an action
functional of fields $a_{\vphantom{k}}^{*}$ and $ a,$
$A(a_{\vphantom{k}}^{*}, a)$ is a function of $a_{\vphantom{k}}^{*}$
and $ a$, and $\mathscr{D}(a_{\vphantom{k}}^{*}, a)$ denotes a
formal measure on a space of fields $a_{\vphantom{k}}^{*}$ and $a.$
All problems of quantum field theory are thus reduced to problems of
finding a correct definition and a computation method of the
functional integrals (\ref{e:fundamental}). However, the results of
the integration of (\ref{e:fundamental}) are not unique and depend
on the chosen method for carrying out the computation of these
functional integrals. This is why in their monograph
\cite{KobzarevManin} the mathematicians Kobzarev and Manin  have
expressed the following statement about (\ref{e:fundamental}):
``From a mathematician's viewpoint almost every such computation is
in fact a half-baked and \textit{ad hoc} definition, but a readiness
to work heuristically with such a priori undefined expressions as
(\ref{e:fundamental}) is a must in this domain."
The most standard method for the integrations of
(\ref{e:fundamental}) is their reduction to Gaussian integrals,
whose theory is the only developed chapter of infinite-dimensional
integration, and then to use an appropriate perturbation expansion.
In this method one divides the action functional
$S(a_{\vphantom{k}}^{*}, a)$ into a sum of two terms
\begin{equation}\label{e:S=S0+S1}
    S(a_{\vphantom{k}}^{*}, a)=S_{0}(a_{\vphantom{k}}^{*}, a)+S_{I}(a_{\vphantom{k}}^{*}, a),
\end{equation}
where $S_{0}$ has a bilinear form in the fields
$a_{\vphantom{k}}^{*}$ and $ a.$ Thus the corresponding partition
function
\begin{equation}\label{e:Z0}
    \mathscr{Z}_{0}=\int e^{-S_{0}(a_{\vphantom{k}}^{*}, a)}\mathscr{D}(a_{\vphantom{k}}^{*}, a)
\end{equation}
is a Gaussian integral and can be exactly and explicitly evaluated.
The total partition function $\mathscr{Z}$ can be expressed in the
form
\begin{equation}\label{e:Z=Z0 stred exp(S1)}
    \mathscr{Z}=\mathscr{Z}_{0}\big\langle {\exp(-S_{I})}\big\rangle_{0},
\end{equation}
where
\begin{equation}\label{e:stred exp(S1)}
    \big\langle {\exp(-S_{I})}\big\rangle_{0}=\sum_{\nu=0}^{\infty}\frac{(-1)^{\nu}}{\nu
    !}\langle {S_{I}^{\nu}}\rangle_{0}
\end{equation}
is the infinite perturbation series involving quantum expectation
values of $\langle {S_{I}^{\nu}}\rangle_{0}$ evaluated with respect
to the chosen action $S_{0}.$ The separation (\ref{e:S=S0+S1}) of
the action functional $S(a_{\vphantom{k}}^{*}, a)$ as the sum of
terms $S_{0}$ and $S_{I}$ is, however, not unique. There are
infinitely many ways to select $S_{0}.$ To each selected $S_{0}$ one
obtains a different result for the perturbation series (\ref{e:stred
exp(S1)}). Thus the results  of functional integrations of
(\ref{e:functional Z}) as expressed by (\ref{e:Z=Z0 stred exp(S1)})
seem to be indeed as ``half-baked and \textit{ad hoc} definitions"
from the mathematician's viewpoint.

If quantum field theory based on the functional integrals
(\ref{e:fundamental}) is equivalent to the same quantum field
theory formulated in the operator formalism, then the relations

\begin{subequations}\label{e:equivalence 1.7 and 1.8}
\begin{equation}\label{e:equivalence 1.7}
  \int e^{-S(a_{\vphantom{k}}^{*}, a)}\mathscr{D}(a_{\vphantom{k}}^{*}, a)
  =
  \mathop{\rm Tr} e^{-\beta H(\bm{a}_{\vphantom{k}}^{+},\bm{a})}
\end{equation}
and
\begin{eqnarray}\label{e:equivalence 1.8}
  &&\phantom{=}\int
  A(a_{\vphantom{k}}^{*}, a)e^{-S(a_{\vphantom{k}}^{*}, a)}\mathscr{D}(a_{\vphantom{k}}^{*}, a)\nonumber\\
  &&=
  \mathop{\rm Tr}
  \Big\{
    A(\bm{a}_{\vphantom{k}}^{+},\bm{a})e^{-\beta H(\bm{a}_{\vphantom{k}}^{+},\bm{a})}
  \Big\}
\end{eqnarray}
\end{subequations}
must be satisfied, where $H(\bm{a}_{\vphantom{k}}^{+},\bm{a})$ is
the Hamiltonian operator corresponding to the action
$S(a_{\vphantom{k}}^{*}, a)$, the operator
$A(\bm{a}_{\vphantom{k}}^{+},\bm{a})$ corresponds to its normal
symbol $A(a_{\vphantom{k}}^{*}, a)$, $\bm{a}_{\vphantom{k}}^{+}$ and
$\bm{a}$ are field operators, and $\beta=1/T$ is the inverse
temperature.

From the last relations it follows that the functional integrals
(\ref{e:fundamental}) cannot be defined in such a way as to give a
unique result after a process of their integrations because the
right-hand sides of the relations (\ref{e:equivalence 1.7 and 1.8})
are distinct for each inequivalent representation of the commutator
or anticommutator ring of field operators
$\bm{a}_{\vphantom{k}}^{+}$ and $\bm{a}$ entering the Hamiltonian
$H(\bm{a}_{\vphantom{k}}^{+},\bm{a})$ and the operator
$A(\bm{a}_{\vphantom{k}}^{+},\bm{a}).$ The reasons are given by the
following arguments.

In the operator approach one assumes to have a complete set of
annihilation and creation operators
$\bm{a}_{\bm{k},\sigma}^{\vphantom{+}}$ and
$\bm{a}_{\bm{k},\sigma}^{+}$ of particles in quantum states denoted
by quantum numbers $(\bm{k},\sigma),$ as, for example, by the
momentum $\bm{k}$ and the spin $\sigma.$ In any quantum field theory
the number of the pairs of the operators
$(\bm{a}_{\bm{k},\sigma}^{\vphantom{+}},\bm{a}_{\bm{k},\sigma}^{+})$
is infinite. These operators satisfy the canonical commutation or
anticommutation relations
\begin{subequations}\label{e:ring}
\begin{eqnarray}
\label{e:a,a+}
    \{\bm{a}_{\bm{k},\sigma}^{\vphantom{+}},\bm{a}_{\bm{k}',\sigma'}^{+}\}
    &=&
    \delta_{\bm{k}\bm{k}'}\delta_{\sigma\sigma'}, \\
\label{e:aa, a+a+}
    \{\bm{a}_{\bm{k},\sigma}^{\vphantom{+}},\bm{a}_{\bm{k}',\sigma'}^{\vphantom{+}}\}
    &=&
    \{\bm{a}_{\bm{k},\sigma}^{+},\bm{a}_{\bm{k}',\sigma'}^{+}\}=0.
\end{eqnarray}
\end{subequations}
The operators act on state vectors $\Psi$ which span a Hilbert space
$\mathscr{H}.$ In order to achieve a unique specification of the
commutator or anticommutator ring of the operators (\ref{e:ring}),
one requires in addition to (\ref{e:ring}) the existence of a vacuum
state $\Phi^{\vphantom{+}}_{0}$ for which
\begin{equation}\label{e:vacuum condition}
    \bm{a}_{\bm{k},\sigma}^{\vphantom{+}}\Phi^{\vphantom{+}}_{0}=0
\end{equation}
for all $(\bm{k},\sigma).$

In this case, the Hilbert space $\mathscr{H}$ is the space for a
representation of the commutator or anticommutator ring of the
operators (\ref{e:ring}) with the auxiliary condition (\ref{e:vacuum
condition}). Since the operators
$\bm{a}_{\bm{k},\sigma}^{\vphantom{+}}$ and
$\bm{a}_{\bm{k},\sigma}^{+}$ form a complete set of operators, a
Hamiltonian $\bm{H}$ of a given system can be expressed as a given
function of $\bm{a}_{\bm{k},\sigma}^{\vphantom{+}}$ and
$\bm{a}_{\bm{k},\sigma}^{+},$ i.e.
\begin{equation}\label{e:Hamiltonian}
    \bm{H}=H(\bm{a}_{\vphantom{k}}^{+},\bm{a}).
\end{equation}
The partition function $\mathscr{Z}$ of the system is expressed as
the trace of the density matrix
\begin{equation}\label{e:rho}
    \bm{\rho}=e^{-\beta\bm{H}},
\end{equation}
i.e.,
\begin{subequations}\label{e:Z <A> - operator}
\begin{equation}\label{e:Z s trace}
    \mathscr{Z}=\mathop{\rm Tr} e^{-\beta\bm{H}},
\end{equation}
and the statistical average values corresponding to physical
observables associated with the operators
$A(\bm{a}_{\vphantom{k}}^{+},\bm{a})$ are given by
\begin{equation}\label{e:average value of A}
    \langle {A}\rangle=\frac{1}{\mathscr{Z}}\mathop{\rm Tr}\big\{A(\bm{a}_{\vphantom{k}}^{+},\bm{a})e^{-\beta\bm{H}}\big\}.
\end{equation}
\end{subequations}

As long as the number of the operators
$\bm{a}_{\bm{k},\sigma}^{\vphantom{+}},$
$\bm{a}_{\bm{k},\sigma}^{+}$ entering the commutator or
anticommutator ring (\ref{e:ring}) is finite, there is only one
inequivalent
representation of the relations (\ref{e:ring}) and (\ref{e:vacuum condition}). %
However, in quantum field theories describing systems with an
infinite number of degrees of freedom, the algebraic structure
(\ref{e:ring}) has infinitely many inequivalent representations
\cite{Haag}. Intuitively speaking, one can say that there exists
infinitely many different matrix realizations of the operators
$\bm{a}_{\bm{k},\sigma}^{\vphantom{+}}$ and
$\bm{a}_{\bm{k},\sigma}^{+}$ satisfying the same algebraic structure
(\ref{e:ring}). The situation reminds us distantly of a Lie algebra
of a non-compact group which has infinitely many unitary irreducible
representations realized by sets of infinite-dimensional matrices.
Thus, to each inequivalent representation of the commutator or
anticommutator ring (\ref{e:ring}) one has to associate the
corresponding representation of the Hamiltonian
(\ref{e:Hamiltonian}) and the density matrix (\ref{e:rho}).

Intuitively speaking, one can say that the matrix form of the
Hamiltonian (\ref{e:Hamiltonian}) and the density matrix
(\ref{e:rho}) are distinct for each inequivalent representation of
(\ref{e:ring}). This implies that the partition function
$\mathscr{Z}$ and the average values $\langle {A}\rangle$ as given
by (\ref{e:fundamental}) can give rise to various results depending
on the chosen inequivalent representations of the ring
(\ref{e:ring}). This non-uniqueness of the value of the partition
function $\mathscr{Z}$ and of the average values $\langle
{A}\rangle$ associated with the given Hamiltonian present in the
operator formalism (\ref{e:ring})-(\ref{e:Z <A> - operator}) should
be preserved in quantum field theory based on the functional
integrals (\ref{e:fundamental}) if these two approaches are
equivalent. They should be equivalent because the integrand
$\exp\{-S(a_{\vphantom{k}}^{*}, a)\}$ entering the functional
integral (\ref{e:functional Z}) is in fact the kernel of the density
matrix $\bm{\rho}=\exp\{-\beta\bm{H}\}$
\cite{Berezin,FaddeevSlavnov,Popov,Vasiliev,ChaichianDemichev}. For
these reasons, the functional integral (\ref{e:functional Z}) or
(\ref{e:<A>}) can be as well defined as is permitted by a freedom
present in the operator approach to quantum field theory. This
freedom is associated with the existence of the inequivalent
representations of the commutator or anticommutator ring
(\ref{e:ring}) of field operators.

For a detailed understanding  of the fact that the computation of
the functional integrals (\ref{e:fundamental}) can give rise to
various different results, we will analyze the process of their
integrations from three different aspects.  Firstly, we analyze a
certain class of inequivalent representations of the anticommutator
ring (\ref{e:ring}) and show that to each inequivalent
representation one has to associate a distinct action functional
$S(a_{\vphantom{k}}^{*}, a).$ Thus, even the explicit form of the
action functional $S(a_{\vphantom{k}}^{*}, a)$ corresponding to a
given Hamiltonian $\bm{H}$ is not unique, but depends on the chosen
inequivalent representation of the anticommutator ring of field
operators. Secondly, we study perturbation series in both the
operator and functional integral formalisms of quantum field
theories in order to show that by selecting an unperturbed part of a
Hamiltonian and the corresponding unperturbed action functional, one
in fact selects one inequivalent representation of the commutator or
anticommutator ring (\ref{e:ring}) of field operators.  Thirdly, we
demonstrate explicitly the properties of the functional integrals
mentioned above on a toy model of quantum field theory. We construct
a simple action functional $S(a_{\vphantom{k}}^{*}, a)$ which
permits  the exact evaluation of the functional integral
(\ref{e:functional Z}), but nonetheless it leads to infinitely many
different results corresponding to infinitely many perturbation
series.

\section{\label{s:inequivalent representations} Inequivalent representations}
For the sake of simplicity, we start by considering a complete set
of annihilation and creation field operators
$\bm{a}_{k,\sigma}^{\vphantom{+}}$ and $\bm{a}_{k,\sigma}^{+}$ of
the fermion type. Let the index $k$ run over integer numbers over
the interval $k\in[-\frac{N}{2},\frac{N}{2}]$ and $\sigma$ denote
spin $\frac{1}{2}$ projection of a fermion, i.e.,
$$\sigma=\uparrow,\downarrow = +,-.$$
In order to have a quantum field theory, we take the limit
$N\rightarrow\infty.$ The field operators
$\bm{a}_{k,\sigma}^{\vphantom{+}}$ and $\bm{a}_{k,\sigma}^{+}$
satisfy the anticommutator ring
\begin{subequations}\label{e:ring-f}
\begin{eqnarray}
\label{e:fermionic aa+}
    \{\bm{a}_{k,\sigma}^{\vphantom{+}},\bm{a}_{k',\sigma'}^{+}\}
    &=&\delta_{kk'}\delta_{\sigma\sigma'}\\
\label{e:fermionic aa, a+a+}
    \{\bm{a}_{k,\sigma}^{\vphantom{+}},\bm{a}_{k',\sigma'}^{\vphantom{+}}\}
    &=&
    \{\bm{a}_{k,\sigma}^{+},\bm{a}_{k',\sigma'}^{+}\}=0
\end{eqnarray}
\end{subequations}
with the subsidiary condition
\begin{equation}\label{e:vacuum-f}
    \bm{a}_{k,\sigma}^{\vphantom{+}}\Phi^{\vphantom{+}}_{0}=0
\end{equation}
on the vacuum state $\Phi_{0}$ for all $(k,\sigma).$ The
representation space for the anticommutator ring (\ref{e:ring-f})
with the subsidiary condition (\ref{e:vacuum-f}) can be chosen to be
the Hilbert space $\mathscr{H}$ spanned by the basis vectors
${\Psi^{\vphantom{+}}}_{\{n_{k,\sigma}\}}$ defined by the formula
\begin{equation}\label{e:basis-f}
    {\Psi^{\vphantom{+}}}_{\{n_{k,\sigma}\}}
    \stackrel{\text{def}}{=}
    \lim_{N\rightarrow\infty}\prod_{k,\sigma}\big(\bm{a}_{k,\sigma}^{+}\big)^{n_{k\,\sigma}}\Phi^{\vphantom{+}}_{0},
\end{equation}
where $n_{k,\sigma}=0,1$ are the occupation numbers of fermions in
states $(k,\sigma)$, and $\{n_{k,\sigma}\}$ denotes an array with an
infinite number of items $0$ and $1.$ Each such infinite array
$\{n_{k,\sigma}\}$ specifies one of the basis vectors of the Hilbert
space $\mathscr{H}.$

The Hamiltonian $\bm{H}$ governing a physical system in a quantum
field theory is a function of the operators
$\bm{a}_{k,\sigma}^{\vphantom{+}}$ and $\bm{a}_{k,\sigma}^{+}.$ Let
its normal form be denoted by
\begin{equation}\label{e:H-normal form}
    \bm{H}=H(\bm{a}_{\vphantom{k}}^{+},\bm{a}).
\end{equation}
To the normal form of $H(\bm{a}_{\vphantom{k}}^{+},\bm{a})$ one
assigns the normal symbol
\begin{equation}\label{e:H-normal symbol}
    H=H\big(a_{\vphantom{k}}^{*}(\tau), a(\tau)\big)
\end{equation}
of the operator (\ref{e:H-normal form}) in which every operator
$\bm{a}_{k,\sigma}^{\vphantom{+}}$ and $\bm{a}_{k,\sigma}^{+}$ is
replaced by the Grassmann generators
$a_{k,\sigma}^{\vphantom{*}}(\tau)$ and $a_{k,\sigma}^{*}(\tau)$,
respectively
\cite{Berezin,FaddeevSlavnov,Popov,Vasiliev,ChaichianDemichev}.
These generators are thus enumerated also by the continuous
parameter $\tau$ in addition to the quantum numbers $(k,\sigma)$.
The action functional $S=S(a_{\vphantom{k}}^{*}, a)$ is then defined
by the integral
\begin{eqnarray}\label{e:S(a*,a)}
    S(a_{\vphantom{k}}^{*}, a)
    &=&
    \int_{0}^{\beta}{{\rm d}\tau\mkern 6mu}\lim_{N\rightarrow\infty}
    \bigg\{
        \sum_{k,\sigma}a_{k,\sigma}^{*}(\tau)\dot{a}_{k,\sigma}^{\vphantom{*}}(\tau)\nonumber\\
        &&+\ %
        H\big(a_{\vphantom{k}}^{*}(\tau), a(\tau)\big)
    \bigg\},
\end{eqnarray}
where $\dot{a}_{k,\sigma}^{\vphantom{*}}(\tau)$ denotes the ``time"
derivative of $a_{k,\sigma}^{\vphantom{*}}(\tau).$

The partition function $\mathscr{Z}$ of the system is expressed as
\begin{equation}\label{e:Z-a}
    \mathscr{Z}=\mathop{\rm Tr} e^{-\beta H(\bm{a}_{\vphantom{k}}^{+},\bm{a})}
\end{equation}
in the operator formalism, or as the integral
\begin{equation}\label{e:Z-Grassmann}
    \mathscr{Z}=\int e^{-S(a_{\vphantom{k}}^{*}, a)}\mathscr{D}(a_{\vphantom{k}}^{*}, a)
\end{equation}
in the functional integral formalism of the quantum field theory.
From the relations (\ref{e:Z-a}) and (\ref{e:Z-Grassmann}), one
evidently sees that $\exp\big\{-S(a_{\vphantom{k}}^{*}, a)\big\}$ is
in fact the kernel of the density matrix $\exp\big\{-\beta
H(\bm{a}_{\vphantom{k}}^{+},\bm{a})\big\}$ operator, and therefore
the expressions (\ref{e:Z-a}) and (\ref{e:Z-Grassmann}) should give
the same result if these two approaches to the quantum field theory
are equivalent.

Next, we study a class of inequivalent representations of the
anticommutator ring (\ref{e:ring-f}). We start from the operators
$\bm{a}_{k,\sigma}^{\vphantom{+}}$ and $\bm{a}_{k,\sigma}^{+}$
obeying (\ref{e:ring-f}) and
introduce the %
``unitary" transformations
\begin{equation}\label{e:c}
    \bm{c}_{k,\sigma}^{\vphantom{+}}=e^{i\bm{Q}}\bm{a}_{k,\sigma}^{\vphantom{+}}e^{-i\bm{Q}},\qquad
    \bm{c}_{k,\sigma}^{+}=e^{i\bm{Q}}\bm{a}_{k,\sigma}^{+}e^{-i\bm{Q}},
\end{equation}
where $\bm{Q}$ is the Hermitian operator
\begin{subequations}\label{e:Q}
\begin{eqnarray}
\label{e:Q=sumT}
        \bm{Q}&=&\lim_{N\rightarrow\infty}\sum_{k}\alpha_{k}\bm{T}_{k},\\
\label{e:T}
        \bm{T}_{k}&=&i(\bm{a}_{k,+}^{+}\bm{a}_{-k,-}^{+}-\bm{a}_{-k,-}^{\vphantom{+}}\bm{a}_{k,+}^{\vphantom{+}}),
\end{eqnarray}
\end{subequations}
and $\alpha_{k}$ are arbitrary real parameters. The anticommutation
relations for the transformed operators
$\bm{c}_{k,\sigma}^{\vphantom{+}}$ and $\bm{c}_{k,\sigma}^{+}$ are,
of course, the same as those given by (\ref{e:ring-f}). The operator
$e^{i\bm{Q}}$ can be expressed as the infinite product
\begin{equation}\label{e:exp iQ}
        e^{i\bm{Q}}
        =
        \lim_{N\rightarrow\infty}\prod_{k=-N/2}^{N/2}
        \big[
                1+i\bm{T}_{k}\sin\alpha_{k}-\bm{T}_{k}^{2}(1-\cos\alpha_{k})
        \big],
\end{equation}
where
\begin{equation}\label{e:T2}
        \bm{T}_{k}^{2}=2\bm{a}_{k,+}^{+}\bm{a}_{k,+}^{\vphantom{+}}\bm{a}_{-k,-}^{+}\bm{a}_{-k,-}^{\vphantom{+}}
        -\bm{a}_{k,+}^{+}\bm{a}_{k,+}^{\vphantom{+}}
        -\bm{a}_{-k,-}^{+}\bm{a}_{-k,-}^{\vphantom{+}}
        +1.
\end{equation}
The transformations (\ref{e:c}), if elaborated with %
(\ref{e:exp iQ}), are similar to the well-known Bogoliubov-Valatin
transformations \cite{Bogoliubov,Valatin}
\begin{subequations}\label{e:c-bogoljubov}
\begin{eqnarray}
\label{e:cm+-bogoljubov}
    \bm{c}_{k,+}^{\vphantom{+}}&=&u_{k}\bm{a}_{k,+}^{\vphantom{+}}+v_{k}\bm{a}_{-k,-}^{+},\\
\label{e:cm--bogoljubov}
    \bm{c}_{k,-}^{\vphantom{+}}&=&u_{k}\bm{a}_{k,-}^{\vphantom{+}}-v_{k}\bm{a}_{-k,+}^{+},\\
\label{e:cp+-bogoljubov}
    \bm{c}_{k,+}^{+}&=&u_{k}\bm{a}_{k,+}^{+}+v_{k}\bm{a}_{-k,-}^{\vphantom{+}},\\
\label{e:cp--bogoljubov}
    \bm{c}_{k,-}^{+}&=&u_{k}\bm{a}_{k,-}^{+}-v_{k}\bm{a}_{-k,+}^{\vphantom{+}},
\end{eqnarray}
where
\begin{equation}\label{e:uk,vk}
u_{k}=\cos{\alpha_{k}}\qquad \text{and}\qquad v_{k}=\sin
\alpha_{k}.
\end{equation}
\end{subequations}
In the limit $N\rightarrow\infty$, the operator $\bm{Q}$ given by
(\ref{e:Q}) is not a proper operator, but transforms every vector
$\Psi$ of the Hilbert space $\mathscr{H}$ into
$\Psi'=e^{i\bm{Q}}\Psi$ of the Hilbert space $\mathscr{H}'$ with
unexpected properties.

Let us denote by $\varphi_{\{n_{k}\}}$ any basis vector of
$\mathscr{H}$ given by the formula
\begin{equation}\label{e:basis-varphi}
        \varphi_{\{n_{k}\}}
        =
        \lim_{N\rightarrow\infty}\prod_{k=-N/2}^{N/2}
        \big(
                \bm{a}_{k,+}^{+}\bm{a}_{-k,-}^{+}
        \big)^{n_{k}}\Phi^{\vphantom{+}}_{0},
\end{equation}
where $n_{k}=n_{k,+}=n_{-k,-}=0,1.$ All the basis vectors
$\varphi_{\{n_{k}\}}$ form a subspace of $\mathscr{H}.$ The
transformation $e^{i\bm{Q}}$ transforms every basis vector
$\varphi_{\{n_{k}\}}$ into one
$\varphi'_{\{n_{k}\}}=e^{i\bm{Q}}\varphi_{\{n_{k}\}}$ of
$\mathscr{H}'$, given by the formula
\begin{eqnarray}
    \varphi'_{\{n_{k}\}}
    &=&
        \lim_{N\rightarrow\infty}\prod_{k=-N/2}^{N/2}
        \Big\{
                \big[
                        \delta_{n_{k},1}-(\bm{a}_{k,+}^{+}\bm{a}_{-k,-}^{+})^{n_{k}+1}
                \big]\sin\alpha_{k}
                        \nonumber\\
    &&+\ %
                        (\bm{a}_{k,+}^{+}\bm{a}_{-k,-}^{+})^{n_{k}}\cos\alpha_{k}
        \Big\}\Phi^{\vphantom{+}}_{0}.
\end{eqnarray}
The result is that the scalar product
$\big({\Psi^{\vphantom{+}}},e^{i\bm{Q}}\varphi_{\{n_{k}\}}\big)$ of
every basis vector $\Psi$ given by (\ref{e:basis-f}) of
$\mathscr{H}$ is either identically equal to zero or equal to the
infinite product
\begin{subequations}\label{e:skalar Psi,varphi}
\begin{equation}\label{e:skalar Psi,varphi-a}
  \big({\Psi^{\vphantom{+}}}_{\{n'_{k'}\}},e^{i\bm{Q}}\varphi_{\{n\vphantom{'}_{k\vphantom{'}}\}}\big)
  =
  \lim_{N\rightarrow\infty}
  \prod_{k,k'}
  \Big(
    D_{s}\sin\alpha_{k}+D_{c}\cos\alpha_{k}
  \Big),
\end{equation}
where
\begin{eqnarray}
\label{e:skalar Psi,varphi-b}
  D_{s}
  &=&
%
%
  \delta_{1,n_{k}}\delta_{0,n'_{k'}}
%
%
  -
  \delta_{0,n\vphantom{'}_{k\vphantom{'}}}\delta_{1,n'_{k'}}\delta_{k,k'},\\
\label{e:skalar Psi,varphi-c}
  D_{c}
  &=&
  \delta_{n\vphantom{'}_{k\vphantom{'}},n'_{k'}}\delta_{k,k'},
\end{eqnarray}
\end{subequations}
which also diverges to zero in the limit $N\rightarrow\infty$ for
any suitable parameters $\alpha_{k}.$ Thus the Hilbert space
$\mathscr{H}'$ contains a subspace of the state vectors
$\varphi'_{\{n_{k}\}}$ which are orthogonal to every vector $\Psi$
of $\mathscr{H}.$ To make the conclusion as in Haag's paper
\cite{Haag}, $\bm{c}_{k,\sigma}^{\vphantom{+}}$ and
$\bm{c}_{k,\sigma}^{+}$ given by (\ref{e:c}) are operators
satisfying the same canonical anticommutator ring as
(\ref{e:ring-f}), i.e.,
\begin{subequations}\label{e:ring-c}
\begin{eqnarray}
\label{e:fermionic cc+}
    \{\bm{c}_{k,\sigma}^{\vphantom{+}},\bm{c}_{k',\sigma'}^{+}\}
    &=&\delta_{kk'}\delta_{\sigma\sigma'}\\
\label{e:fermionic cc, c+c+}
    \{\bm{c}_{k,\sigma}^{\vphantom{+}},\bm{c}_{k',\sigma'}^{\vphantom{+}}\}
    &=&
    \{\bm{c}_{k,\sigma}^{+},\bm{c}_{k',\sigma'}^{+}\}=0,
\end{eqnarray}
\end{subequations}
but there is no proper unitary transformation connecting these two
operator systems. In other words, they belong to inequivalent
representations of the same anticommutator ring (\ref{e:ring-f})
or (\ref{e:ring-c}) of the field operators. Each inequivalent
representation of (\ref{e:ring-c}) is specified by the chosen
infinite set of parameters $\alpha_{k}$ entering the
transformations (\ref{e:c})-(\ref{e:c-bogoljubov}). Thus, the
number of the inequivalent representations of (\ref{e:ring-c}) is
infinite.

For each inequivalent representation of (\ref{e:ring-c}), we can
construct the transformed Hamiltonian
\begin{equation}\label{e:hamiltonian-c}
        \tilde{H}(\bm{c}_{\vphantom{k}}^{+},\bm{c})
        =
        e^{i\bm{Q}}H(\bm{a}_{\vphantom{k}}^{+},\bm{a})e^{-i\bm{Q}}
\end{equation}
in its normal form by employing the anticommutation relations
(\ref{e:ring-c}). The corresponding partition function
\begin{equation}\label{e:Z-c}
        \tilde{\mathscr{Z}}=\mathop{\rm Tr} e^{-\beta\tilde{H}(\bm{c}_{\vphantom{k}}^{+},\bm{c})}
\end{equation}
can be different from that given by (\ref{e:Z-a}) because the
operators $H(\bm{a}_{\vphantom{k}}^{+},\bm{a})$ and
$\tilde{H}(\bm{c}_{\vphantom{k}}^{+},\bm{c})$ act in different
Hilbert spaces $\mathscr{H}$ and $\mathscr{H}'$, respectively.

In the functional formalism of quantum field theory, we construct
the normal symbol $\tilde{H}( c_{\vphantom{k}}^{*}(\tau), c(\tau))$
of the operator $\tilde{H}(\bm{c}_{\vphantom{k}}^{+},\bm{c})$ and
the action functional
\begin{eqnarray}\label{e:S-c}
    \tilde{S}( c_{\vphantom{k}}^{*}, c)
    &=&
    \int_{0}^{\beta}{{\rm d}\tau\mkern 6mu}
    \bigg\{
        \lim_{N\rightarrow
        \infty}\sum_{k,\sigma} c_{k,\sigma}^{*}(\tau)\dot{c}_{k,\sigma}^{\vphantom{*}}(\tau)\nonumber\\
    &&+\ %
        \tilde{H}\big( c_{\vphantom{k}}^{*}(\tau), c(\tau)\big)
    \bigg\}
\end{eqnarray}
for each inequivalent representation of the anticommutator ring
(\ref{e:ring-c}) of the field operators
$\bm{c}_{k,\sigma}^{\vphantom{+}}$ and $\bm{c}_{k,\sigma}^{+}.$ The
corresponding partition function $\tilde{\mathscr{Z}}$ in the
functional integral form
\begin{equation}\label{e:Z-ctilde}
        \tilde{\mathscr{Z}}=\int e^{-\tilde{S}( c_{\vphantom{k}}^{*}, c)}\mathscr{D}( c_{\vphantom{k}}^{*}, c)
\end{equation}
gives the distinct result for each inequivalent representation.
One can be easily convinced by studies of concrete Hamiltonians
with interactions between fields that the functional integral
(\ref{e:Z-ctilde}) cannot be transformed into that of
(\ref{e:Z-Grassmann}) by the transformations
(\ref{e:c-bogoljubov}),
\begin{subequations}\label{e:c-bg}
\begin{eqnarray}
\label{e:cm+-bg}
     c_{k,+}^{\vphantom{*}}(\tau)&=&u_{k} a_{k,+}^{\vphantom{*}}(\tau)+v_{k}a_{-k,-}^{*}(\tau),\\
\label{e:cm--bg}
     c_{k,-}^{\vphantom{*}}(\tau)&=&u_{k} a_{k,-}^{\vphantom{*}}(\tau)-v_{k}a_{-k,+}^{*}(\tau),\\
\label{e:cp+-bg}
     c_{k,+}^{*}(\tau)&=&u_{k}a_{k,+}^{*}(\tau)+v_{k} a_{-k,-}^{\vphantom{*}}(\tau),\\
\label{e:cp--bg}
     c_{k,-}^{*}(\tau)&=&u_{k}a_{k,-}^{*}(\tau)-v_{k} a_{-k,+}^{\vphantom{*}}(\tau),
\end{eqnarray}
\end{subequations}
of the integration variables.

We conclude this section by stating that a given Hamiltonian
$\bm{H}$ of a system leads to distinct results for the partition
function $\mathscr{Z}$, and for statistical average values $\langle
{A}\rangle$ depending on the chosen inequivalent representation of
the anticommutator ring of field operators in both the operator and
the functional integral approach to quantum field theory. Our
analysis of inequivalent representations of the anticommutator ring
of field operators (\ref{e:ring-f}) presented above can be
generalized in a straightforward way to any set of quantum numbers
$(k,\sigma)$ and to many other classes of inequivalent
representations.

\section{\label{s:perturbation} Perturbation series}
We begin with an elucidation of how one tacitly selects a single
inequivalent representation of the commutator or anticommutator ring
(\ref{e:ring}) of field operators in a practical application of
quantum field theory. In quantum field theories with interactions
between fields, there is not known even one physical example with an
exact solution. In all practical calculations, one divides the
Hamiltonian $\bm{H}$ of a system into the sum
\begin{equation}\label{e:pert:Hamiltonian 3.1}
    \bm{H}=H_{0}(\bm{a}_{\vphantom{k}}^{+},\bm{a})+H_{\text{I}}(\bm{a}_{\vphantom{k}}^{+},\bm{a}),
\end{equation}
where $H_{0}$ is called the unperturbed Hamiltonian, and the
remaining term $H_{I}$ is called the perturbative part. The
unperturbed Hamiltonian $H_{0}$ is chosen in such a way in order to
be exactly diagonalized, and by this fact  its effects are treated
exactly. Its eigenstates ${\Psi^{\vphantom{+}}}_{\mu},$ where $\mu$
denotes an array with an infinite series of items, form a complete
basis of a Hilbert space $\mathscr{H}.$ This is the representation
space for a single representation of the commutator or
anticommutator ring (\ref{e:ring}) of field operators
$\bm{a}_{k,\sigma}^{\vphantom{+}}$ and $\bm{a}_{k,\sigma}^{+}$
entering the Hamiltonian $\bm{H}.$ The partition function
\begin{equation}\label{e:pert:Z0 3.2}
    \mathscr{Z}_{0}=\mathop{\rm Tr} e^{-\beta H_{0}(\bm{a}_{\vphantom{k}}^{+},\bm{a})}
\end{equation}
can be exactly evaluated and is typical for the chosen
representation. The total partition function $\mathscr{Z}$ is
expressed as the perturbation series
\begin{eqnarray}\label{e:pert:Z-series 3.3}
    \mathscr{Z} &=& \mathop{\rm Tr} e^{-\beta H}=\mathscr{Z}_{0}
    \bigg\langle
        \text{T}\exp\bigg\{-\int_{0}^{\beta}{{\rm d}\tau\mkern 6mu} V(\tau)\bigg\}
    \bigg\rangle_{0}\nonumber \\
    &=& \mathscr{Z}_{0}\sum_{\nu=0}^{\infty}\frac{(-1)^{\nu}}{\nu!}
    \bigg\langle
        \text{T}
    \bigg(\int_{0}^{\beta}{{\rm d}\tau\mkern 6mu} V(\tau)\bigg)^{\nu}
    \bigg\rangle_{0},
\end{eqnarray}
where the symbol $\text{T}$ stands for the time- or
temperature-ordered product and
\begin{equation}\label{e:pert:V tau 3.4}
    V(\tau)=e^{\tau H_{0}}H_{I}e^{-\tau H_{0}}.
\end{equation}
The individual terms of the perturbation series
(\ref{e:pert:Z-series 3.3}) are in one-to-one correspondence with
the perturbation series (\ref{e:Z=Z0 stred exp(S1)}) and
(\ref{e:stred exp(S1)}) in the functional integral method. Namely,
$\exp\{-S_{0}(a_{\vphantom{k}}^{*}, a)\}$ in (\ref{e:Z0}) is the
kernel of the unperturbed density matrix $\bm{\rho}_{0}=\exp\{-\beta
H_{0}(\bm{a}_{\vphantom{k}}^{+},\bm{a})\}$ operator. The statistical
average values $\big\langle\text{T}\big(\int_{0}^{\beta}{{\rm
d}\tau\mkern 6mu} V(\tau)\big)^{\nu}\big\rangle_{0}$ evaluated in
the chosen single inequivalent representation correspond to the
functional integrals $\big\langle S_{I}^{\nu}(a_{\vphantom{k}}^{*},
a)\big\rangle_{0}$ in (\ref{e:stred exp(S1)}). Thus, the
perturbation series (\ref{e:Z=Z0 stred exp(S1)}), (\ref{e:stred
exp(S1)}) and (\ref{e:pert:Z-series 3.3}) in both the functional
integral and the operator formalism of quantum field theory should
give the same results in the chosen inequivalent representation of
the commutator or anticommutator ring of field operators.

However, the splitting of the total Hamiltonian
$H(\bm{a}_{\vphantom{k}}^{+},\bm{a})$ as given by
(\ref{e:pert:Hamiltonian 3.1}) is not unique. One can equally well
divide the same Hamiltonian as
\begin{equation}\label{e:Hamiltonian tilde 3.5}
        \bm{H}=\tilde{H}_{0}(\bm{a}_{\vphantom{k}}^{+},\bm{a})+\tilde{H}_{I}(\bm{a}_{\vphantom{k}}^{+},\bm{a}),
\end{equation}
where the unperturbed Hamiltonian $\tilde{H}_{0}$ is not related to
$H_{0}$ by any proper unitary transforamtion. In this case, the
eigenstates $\Psi'_{\mu}$ of $\tilde{H}_{0}$ form again a complete
basis of a new Hilbert space $\mathscr{H}'$ for another inequivalent
representation of the commutator or anticommutator  ring
(\ref{e:ring}) of field operators. Thus, in this another
inequivalent representation one gets the partition functions
\begin{equation}\label{e:Z0tilde 3.6}
        \tilde{\mathscr{Z}}_{0}=\mathop{\rm Tr} e^{-\beta\tilde{H}_{0}(\bm{a}_{\vphantom{k}}^{+},\bm{a})}
\end{equation}
and
\begin{equation}\label{e:Ztilde 3.7}
        \tilde{\mathscr{Z}}=\tilde{\mathscr{Z}}_{0}\sum_{\nu=0}^{\infty}\frac{(-1)^{\nu}}{\nu!}
        \bigg\langle {\text{T}\bigg(\int_{0}^{\beta}{{\rm d}\tau\mkern 6mu}\tilde{V}(\tau)\bigg)^{\nu}}\bigg\rangle_{0},
\end{equation}
which are distinct from those given by (\ref{e:pert:Z0 3.2}) and
(\ref{e:pert:Z-series 3.3}) because the operators
$H_{0}(\bm{a}_{\vphantom{k}}^{+},\bm{a})$ and
$\tilde{H}_{0}(\bm{a}_{\vphantom{k}}^{+},\bm{a})$ act in different
Hilbert spaces, $\mathscr{H}$ and $\mathscr{H}'$, respectively.

Two different separations (\ref{e:pert:Hamiltonian 3.1}) and
(\ref{e:Hamiltonian tilde 3.5}) of the same Hamiltonian $\bm{H}$
correspond to two different divisions of the action functional
$S(a_{\vphantom{k}}^{*}, a)$ as given by the formulas
\begin{subequations}\label{s:splitting of S}
\begin{eqnarray}
\label{e:S=S0+S1 3.8}
    S(a_{\vphantom{k}}^{*}, a) &=& S_{0}(a_{\vphantom{k}}^{*}, a)+S_{I}(a_{\vphantom{k}}^{*}, a) \\
\label{e:S=S0+S1 3.9}
    S(a_{\vphantom{k}}^{*}, a) &=& \tilde{S}_{0}(a_{\vphantom{k}}^{*}, a)+\tilde{S}_{I}(a_{\vphantom{k}}^{*}, a)
\end{eqnarray}
\end{subequations}
The corresponding partition functions
\begin{subequations}\label{e:Z0 Z0tilde}
\begin{eqnarray}
\label{e:Z0=exp 3.10}
    \mathscr{Z}_{0} &=& \int\mathscr{D}(a_{\vphantom{k}}^{*}, a)e^{-S_{0}(a_{\vphantom{k}}^{*}, a)}, \\
\label{e:Z0=exp 3.11}
    \tilde{\mathscr{Z}}_{0} &=& \int\mathscr{D}(a_{\vphantom{k}}^{*}, a)e^{-\tilde{S}_{0}(a_{\vphantom{k}}^{*}, a)}
\end{eqnarray}
\end{subequations}
are, of course, different because the functional integrals
(\ref{e:Z0 Z0tilde}) contain different integrands which cannot be
transformed one into another by any substitution of the integration
variables. In the same way, the different results for the
corresponding perturbation series
\begin{subequations}\label{e:Z-series}
\begin{eqnarray}
\label{e:Z-series 3.12}
    \mathscr{Z} &=& \int\mathscr{D}(a_{\vphantom{k}}^{*}, a)e^{-S(a_{\vphantom{k}}^{*}, a)}\nonumber\\
    &=&\mathscr{Z}_{0}\sum_{\nu=0}^{\infty}\frac{(-1)^{\nu}}{\nu!}\big\langle {S_{I}^{\nu}(a_{\vphantom{k}}^{*}, a)}\big\rangle_{0}, \\
\label{e:Z-series 3.13}
    \tilde{\mathscr{Z}} &=&
    \int\mathscr{D}(a_{\vphantom{k}}^{*}, a)e^{-S(a_{\vphantom{k}}^{*}, a)}\nonumber\\
    &=&\tilde{\mathscr{Z}}_{0}\sum_{\nu=0}^{\infty}\frac{(-1)^{\nu}}{\nu!}\big\langle {\tilde{S}_{I}^{\nu}(a_{\vphantom{k}}^{*}, a)}\big\rangle_{0}
\end{eqnarray}
\end{subequations}
can be understood from the viewpoint of inequivalent
representations of the commutator  or anticommutator ring of field
operators. The last two formulae may seem to represent a paradox.
Namely, the result of the integration of the same functional
integral depends on the process of its integration by means of
perturbation series. For this reason, the mathematicians Kobzarev
and Manin regard every such computations of functional integrals
done by physicists as ``half-baked and \textit{ad hoc}
definitions."

However, the different results (\ref{e:Z-series 3.12}) and
(\ref{e:Z-series 3.13}) of the same functional integral are not
due to its \emph{a priori} undefined expression, but are due to
the existence of infinitely many inequivalent representations of
the commutator or anticommutator ring of field operators entering
the quantum field theory.

In the next section we demonstrate the properties (\ref{e:Z-series})
with a toy action functional $S(a_{\vphantom{k}}^{*}, a)$ which
permits exact summation of perturbation series and gives rise to
infinitely many different results.

\section{\label{s:explicit} Two explicitly solvable examples}
For the purpose of demonstrating the conclusions of the two
previous sections, we consider the action functional
\begin{eqnarray}\label{e:SN 4.1}
    S_{N}(a_{\vphantom{k}}^{*}, a)
    &=&
    \sum_{\sigma=\pm}\mathop{\mathord{\sum}'}_{k
    }
    \varepsilon a_{k,\sigma}^{*}a_{k,\sigma}^{\vphantom{*}}\nonumber\\
    &&\ %
    -\frac{g}{N}\mathop{\mathord{\sum}'}_{k,k'
    }a_{k,+}^{*}a_{-k,-}^{*} a_{-k',-}^{\vphantom{*}} a_{k',+}^{\vphantom{*}},
\end{eqnarray}
where $\varepsilon$ and $g$ are real parameters, $N$ is a given even
integer number, $k$ and $k'$ are integers $k,k'\in
\big[-\frac{N}{2},\frac{N}{2}\big]$ with the exclusion of $k=k'=0$
which is indicated by the prime on the summation symbols
(${\mathop{\mathord{\sum}'}}$), and $a_{k,\sigma}^{\vphantom{*}},
a_{k,\sigma}^{*}$ are Grassmann variables. Next, we use the
identity.
\begin{widetext}
\begin{eqnarray}\label{e:identity 4.2}
    S_{N}(a_{\vphantom{k}}^{*}, a)
    &=&
        \sum_{\sigma=\pm}
        \mathop{\mathord{\sum}'}_{k
        }\varepsilon a_{k,\sigma}^{*}a_{k,\sigma}^{\vphantom{*}}
        -
        \frac{1}{\gamma^{2}}
        \bigg(
            \gamma^{2}\Delta^{*}-\gamma^{2}\Delta^{*}-\mathop{\mathord{\sum}'}_{k
            }a_{k,+}^{*}a_{-k,+}^{*}
        \bigg)
        \bigg(
            \gamma^{2}\Delta-\gamma^{2}\Delta-\mathop{\mathord{\sum}'}_{k'
            } a_{-k',-}^{\vphantom{*}} a_{k',-}^{*}
        \bigg)\nonumber\\
    &=&\gamma^{2}\Delta^{*}\Delta
        +
        \sum_{\sigma}\mathop{\mathord{\sum}'}_{k
        }\varepsilon a_{k,\sigma}^{*}a_{k,\sigma}^{\vphantom{*}}
        -
        \mathop{\mathord{\sum}'}_{k
        }
        \big(
            \Delta a_{k,+}^{*}a_{-k,-}^{*}+\Delta^{*} a_{-k,-}^{\vphantom{*}} a_{k,+}^{\vphantom{*}}
        \big)\nonumber\\
    && -\ %
        \frac{1}{\gamma^{2}}
        \bigg(
            \gamma^{2}\Delta^{*}-\mathop{\mathord{\sum}'}_{k
            }a_{k,+}^{*}a_{-k,-}^{*}
        \bigg)
        \bigg(
            \gamma^{2}\Delta-\mathop{\mathord{\sum}'}_{k
            } a_{-k,-}^{\vphantom{*}} a_{k,+}^{\vphantom{*}}
        \bigg),
\end{eqnarray}
\end{widetext}
where
\begin{equation}\label{e:gamma}
  \gamma=\sqrt{\frac{N}{g}}
\end{equation}
and $\Delta$, $\Delta^{*}$ are arbitrary complex numbers. We
separate the action $S_{N}(a_{\vphantom{k}}^{*}, a)$ into a sum of
two terms,
\begin{equation}\label{e:SN=S0N+SIN 4.3}
    S_{N}(a_{\vphantom{k}}^{*}, a)=S_{0,N}(a_{\vphantom{k}}^{*}, a)+S_{I,N}(a_{\vphantom{k}}^{*}, a),
\end{equation}
where the unperturbed action $S_{0N}$ is chosen to be
\begin{subequations}\label{e:S0N SIN}
\begin{eqnarray}\label{e:S0N 4.4}
    S_{0,N} &=& \gamma^{2}\Delta\Delta^{*}
    +
    \mathop{\mathord{\sum}'}_{k}
    \big\{
        \varepsilon(a_{k,+}^{*} a_{k,+}^{\vphantom{*}}+a_{k,-}^{*} a_{k,-}^{\vphantom{*}})\nonumber\\
        &&-\ %
        \Delta a_{k,+}^{*}a_{-k,-}^{*}-\Delta^{*} a_{-k,-}^{\vphantom{*}} a_{k,+}^{\vphantom{*}}
    \big\},
\end{eqnarray}
and
\begin{eqnarray}\label{e:SIN 4.5}
    S_{I,N} &=& -\frac{1}{\gamma^{2}}
    \bigg(
        \gamma^{2}\Delta^{*}-\mathop{\mathord{\sum}'}_{k
        }a_{k,+}^{*}a_{-k,-}^{*}
    \bigg)\nonumber\\
    &&\times\ %
    \bigg(
        \gamma^{2}\Delta-\mathop{\mathord{\sum}'}_{k
        } a_{-k,-}^{\vphantom{*}} a_{k,+}^{\vphantom{*}}
    \bigg)
\end{eqnarray}
\end{subequations}
is the perturbative term which will be treated by the perturbation
series. Thus we have infinitely many unperturbed actions
(\ref{e:S0N 4.4}) enumerated by arbitrary complex numbers $\Delta$
and $\Delta^{*}.$

The corresponding unperturbed partition function $\mathscr{Z}_{0,N}$
can be exactly and explicitly calculated with the result
\begin{subequations}\label{e:partition 4.6}
\begin{eqnarray}
\label{e:partition 4.6a}
    \mathscr{Z}_{0,N}&=&E^{2N}\exp{
    \{
        -\gamma^{2}\Delta\Delta^{*}
    \}},\\
\label{e:partition 4.6b}
    E&=&(\varepsilon^{2}+\Delta\Delta^{*})^{1/2}.
\end{eqnarray}
\end{subequations}
The total partition function $\mathscr{Z}_{N}$ is given by the
infinite perturbation series
\begin{equation}\label{e:Z-series 4.7}
    \mathscr{Z}_{N}=\mathscr{Z}_{0,N}\sum_{\nu=0}^{\infty}\frac{(-1)^{\nu}}{\nu!}\big\langle {S_{I,N}^{\nu}(a_{\vphantom{k}}^{*}, a)}\big\rangle_{0}.
\end{equation}
For the purpose of making an explicit calculation of
$\mathscr{Z}_{N}$ it is convenient to introduce the generating
action functional $\mathscr{S}_{0,N}(a_{\vphantom{k}}^{*}, a;
b_{\vphantom{k}}^{*}, b)$ defined by
\begin{eqnarray}
\label{e:s-generating functional 4.8}
  \mathscr{S}_{0,N}(a_{\vphantom{k}}^{*}, a; b_{\vphantom{k}}^{*}, b)
  &=& S_{0,N}
    +
    \frac{ b_{\vphantom{k}}^{*}}{\gamma}\bigg(\gamma^{2}\Delta-\mathop{\mathord{\sum}'}_{k
    } a_{-k,-}^{\vphantom{*}} a_{k,+}^{\vphantom{*}}\bigg)\nonumber\\
    &&+\ %
    \frac{ b}{\gamma}\bigg(\gamma^{2}\Delta^{*}-\mathop{\mathord{\sum}'}_{k
    }a_{k,+}^{*}a_{-k,-}^{*}\bigg),\nonumber\\
\end{eqnarray}
where $ b_{\vphantom{k}}^{*}$ and $ b$ are two complex variables.
With $\mathscr{S}_{0,N}(a_{\vphantom{k}}^{*}, a;
b_{\vphantom{k}}^{*} b)$ we define the generating partition function
$\mathscr{Z}_{0,N}( b_{\vphantom{k}}^{*}, b)$ by the functional
integral
\begin{equation}\label{e:Z0Nb*b 4.9}
    \mathscr{Z}_{0,N}( b_{\vphantom{k}}^{*}, b)=\int\mathscr{D}(a_{\vphantom{k}}^{*}, a)\exp\big\{-\mathscr{S}_{0,N}(a_{\vphantom{k}}^{*}, a; b_{\vphantom{k}}^{*}, b)\big\},
\end{equation}
which is a Gaussian integral and can be explicitly calculated with
the result
\begin{eqnarray}\label{e:Z0N-result 4.10}
  \mathscr{Z}_{0,N}( b_{\vphantom{k}}^{*}, b) &=&
  \bigg\{
    \varepsilon^{2}
    +
    \bigg(
        \Delta+\frac{1}{\gamma} b
    \bigg)
        \bigg(
        \Delta^{*}+\frac{1}{\gamma} b_{\vphantom{k}}^{*}
    \bigg)
  \bigg\}^{N}\nonumber \\
  &&\times
    \exp
    \{
        -\gamma
        (
            \gamma\Delta^{*}\Delta+ b\Delta^{*}+ b_{\vphantom{k}}^{*}\Delta
        )
    \}.\nonumber\\
    &&
\end{eqnarray}

By introducing the ratio
%
\begin{eqnarray}\label{e:W ratio 4.11}
  W_{N}( b_{\vphantom{k}}^{*}, b) &=&
  \frac{\mathscr{Z}_{0,N}( b_{\vphantom{k}}^{*}, b)}{\mathscr{Z}_{0,N}}\nonumber\\
  &=&
  \bigg\{
    1+\frac{1}{N}\frac{g}{E^{2}}
    (
      \gamma\Delta^{*} b+\gamma\Delta  b_{\vphantom{k}}^{*}+ b_{\vphantom{k}}^{*} b
    )
  \bigg\}^{N}\nonumber\\
  &&\times\exp\{-\gamma(\Delta^{*} b+\Delta b_{\vphantom{k}}^{*})\},
\end{eqnarray}
%
we express the total partition function (\ref{e:Z-series 4.7}) in
the form
\begin{equation}\label{e:ZN 4.12}
    \mathscr{Z}_{N}=\lim_{ b, b_{\vphantom{k}}^{*}\rightarrow 0}
    \mathscr{Z}_{0,N}\exp\bigg\{\frac{\partial^{2}}{\partial{ b\partial b_{\vphantom{k}}^{*}}}\bigg\}W_{N}( b_{\vphantom{k}}^{*}, b).
\end{equation}
Now we are prepared to take the limit $N\rightarrow \infty$ in
order to get the results (\ref{e:partition 4.6})-(\ref{e:ZN 4.12})
for the functional integrals with infinitely many integration
variables. In this limit, the asymptotic formula for the ratio
(\ref{e:W ratio 4.11}) has the form
\begin{eqnarray}
\label{e:W-limit 4.13}
  W_{N}( b_{\vphantom{k}}^{*}, b) &=&
  \exp
  \bigg\{
    \gamma
    \bigg(
        \frac{g}{E^{2}}-1
    \bigg)
    (\Delta b_{\vphantom{k}}^{*}+\Delta^{*} b)\nonumber\\
    &&+\ %
    \frac{g}{E^{2}}\, b_{\vphantom{k}}^{*} b
    -\frac{g}{2E^4}(\Delta b_{\vphantom{k}}^{*}+\Delta^{*} b)^2
    \nonumber\\
    &&+O\big(N^{-1/2}\big)
  \bigg\}.
\end{eqnarray}
The last result is inserted into (\ref{e:ZN 4.12}), and by
interchanging the order of the limits $N\rightarrow\infty$ and $
b\rightarrow 0,$ $ b_{\vphantom{k}}^{*}\rightarrow 0$, we get the
asymptotic formula
\begin{eqnarray}\label{e:ZN limit 4.14}
    Z_{N}&=&E^{2N}
    \exp
    \bigg\{
        \frac{N}{g}\Delta\Delta^{*}
        \bigg[
            \Big(1-\frac{g}{E^{2}}\Big)^{2}\nonumber\\
            &&\hspace{-20pt}\times
            \Big(1-\frac{g}{E^{2}}+2\frac{g}{E^{4}}\Delta\Delta^{*}\Big)^{-1}-1
        \bigg]
        +O(1)
    \bigg\}
\end{eqnarray}
With the restrictive condition for the parameter $\Delta\Delta^{*}$
in the form
$$\Big(1-\frac{g}{E^{2}}\Big)\Big(1-\frac{g}{E^{2}}+2\frac{g}{E^{4}}\Delta\Delta{*}\Big)>0.$$

At the end we define the ``density of the grand canonical
potential" $\Omega$ in the ``thermodynamic limit" as
\begin{equation}\label{e:Omega 4.15}
    \Omega=-\lim_{N\rightarrow\infty}\frac{1}{N}\ln\mathscr{Z}_{N}.
\end{equation}
By using the last formula, we obtain the final result
\begin{eqnarray}\label{e:Omega final 4.16}
    \Omega &=&-\frac{\Delta\Delta^{*}}{g}\,
    \bigg\{
        \Big(1-\frac{g}{E^{2}}\Big)^{2}
        \Big(1-\frac{g}{E^{2}}+2\frac{g}{E^{4}}\Delta\Delta^{*}\Big)^{-1}\nonumber\\
    &&\hspace{42pt}-1
    \bigg\}-2\ln{E},
\end{eqnarray}
which explicitly shows that the functional integral
(\ref{e:functional Z}) with the given action functional (\ref{e:SN
4.1}) gives infinitely many grand canonical potentials $\Omega$
enumerated by arbitrary complex numbers $\Delta$ and $\Delta^{*}.$
This simple exactly solvable example demonstrates explicitly that
functional integrals in quantum field theories cannot be regarded
as Newton-Lebesgue integrals. Different results corresponding to
distinct processes of their integrations of the same functional
integral should not be regarded as \textit{ad hoc} definitions for
\textit{a priori} undefined expressions as (\ref{e:fundamental}).
The distinct results associated with the same functional integral
correspond to the existence of inequivalent representations of the
commutator or anticommutator ring of field operators in the
operator approach to quantum field theory.

For demonstration purposes presented in this section, we have
selected the toy action functional (\ref{e:SN 4.1}) which reminds
us of an oversimplified BCS model of superconductivity
\cite{BardeenCooperSchrieffer}. The method presented in this
chapter can be straightforwardly generalized and applied, e.g., to
the realistic BCS model of superconductivity with the action
functional
\begin{widetext}
\begin{eqnarray}\label{e:BCS 4.17}
  S(a_{\vphantom{k}}^{*}, a) &=& \int_{0}^{\beta}{{\rm d}\tau\mkern 6mu}
  \bigg\{
    \sum_{\bm{k},\sigma}
    \big[
       a_{\bm{k},\sigma}^{*}(\tau)\dot{a}_{\bm{k},\sigma}^{\vphantom{*}}(\tau)
      +\xi_{\bm{k}} a_{\bm{k},\sigma}^{*}(\tau) a_{\bm{k},\sigma}^{\vphantom{*}}(\tau)
    \big]\nonumber\\
  &&-\ %
    \frac{g}{V}\sum_{\bm{k},\bm{k}'}
     a_{\bm{k},+}^{*}(\tau) a_{-\bm{k},-}^{*}(\tau) a_{-\bm{k}',-}^{\vphantom{*}}(\tau) a_{\bm{k}',+}^{\vphantom{*}}(\tau)
    \theta\big(\hbar\omega_{D}-|\xi_{\bm{k}}|\big)
    \theta\big(\hbar\omega_{D}-|\xi_{\bm{k}'}|\big)
  \bigg\},
\end{eqnarray}
\end{widetext}
where $\bm{k}$ is the wave vector, $\sigma$ is the spin
$\frac{1}{2}$ projection of an electron,
$$\xi_{\bm{k}}=\frac{\hbar^{2}\bm{k}^{2}}{2m}-\mu$$
is the kinetic energy of an electron counted from the chemical
potential $\mu$, $\omega_{D}$ is the Debye frequency, $g$ is the
squared electron-phonon coupling constant, and $V$ is the volume
of the system of electrons. By the same steps as presented by the
relations (\ref{e:identity 4.2})-(\ref{e:Omega 4.15}), however,
with more involved calculations,
 one can
derive the exact density for the grand canonical potential
\begin{equation}\label{e:BCS Omega 4.18}
    \Omega(T,\mu)=-\lim_{V\rightarrow\infty}\frac{1}{V\beta}\ln\mathscr{Z}
\end{equation}
in the form
\begin{eqnarray}\label{e:BCS Omega 4.19}
    \Omega(T,\mu,\Delta\Delta^{*})
    &=&
    -\frac{\Delta\Delta^{*}}{g}
    \big[(1-D)^{2}(1-D-2\Delta\Delta^{2}C)^{-1}\nonumber\\
    &&\hspace{-41pt}-1\big]-
    \frac{2}{(2\pi)^{3}\beta}\int{\rm d}^{3}{k}
    \bigg\{
        \ln\bigg[
            2\cosh\frac{\beta E_{\bm{k}}}{2}
        \bigg]-\beta\xi_{\bm{k}}
    \bigg\},\nonumber\\
\end{eqnarray}
where
\begin{eqnarray}\label{e:Ek 4.20}
    E_{\bm{k}}&=&\sqrt{\xi_{\bm{k}}^{2}+\Delta\Delta^{*}}\ %
    \theta\big(\hbar\omega_{D}-|\xi_{\bm{k}}|\big)\nonumber\\
    &&+\ %
    \xi_{\bm{k}}\theta\big(|\xi_{\bm{k}}|-\hbar\omega_{D}\big)
\end{eqnarray}
is the energy spectrum of elementary excitations
\begin{equation}\label{e:D 4.21}
    D=\frac{g}{(2\pi)^{3}}\int\frac{{\rm d}^{3}{k}}{2E_{\bm{k}}}
    \theta\big(\hbar\omega_{D}-|\xi_{\bm{k}}|\big)\tanh\frac{\beta E_{\bm{k}}}{2}
\end{equation}
and
$$C=\frac{\partial D}{\partial(\Delta\Delta^{*})}$$
with the restrictive condition on the parameter $\Delta\Delta^{*}$
in the form
$$(1-D)(1-D-2\Delta\Delta^{*}C)>0.$$

The result (\ref{e:BCS Omega 4.19}) shows again that the
functional integral (\ref{e:functional Z}) with the given action
(\ref{e:BCS 4.17}) gives infinitely many densities of the
grand-canonical potential $\Omega(T,\mu)$ enumerated by arbitrary
complex values $\Delta$ and $\Delta^{*}$ which are called the gap
functions.

The gap functions $\Delta$ and $\Delta^{*}$ determine the parameters
$\alpha_{\bm{k}}$ in the transformations
(\ref{e:c})-(\ref{e:c-bogoljubov}) by the relation
\begin{equation}\label{e:alpha 4.22}
  \sin^{2}\alpha_{\bm{k}}=\frac{1}{2}
  \bigg(
    1-\frac{\xi_{\bm{k}}}{\sqrt{\xi_{\bm{k}}^{2}+\Delta\Delta^{*}}}
  \bigg)
  \theta\big(\hbar\omega_{D}-|\xi_{\bm{k}}|\big).
\end{equation}
The given infinite set of the parameters $\alpha_{k}$ specifies the
corresponding inequivalent representation of the anticommutator ring
(\ref{e:ring-c}) of the field operators $\bm{a}_{\bm{k},\sigma}^{+}$
and $\bm{a}_{\bm{k},\sigma}^{\vphantom{+}}.$ Thus, the values of the
functional integral (\ref{e:functional Z}) with the action
functional (\ref{e:BCS 4.17}) leading to the density (\ref{e:BCS
Omega 4.19}) enumerated by given gap functions $\Delta$ and
$\Delta^{*}$ correspond to distinct inequivalent representations of
the anticommutator ring (\ref{e:ring-c}) of the field operators
$\bm{a}_{\bm{k},\sigma}^{+}$ and
$\bm{a}_{\bm{k},\sigma}^{\vphantom{+}}.$

The second law of thermodynamics, however, requires the density
$\Omega(T,\mu)$ at given values of the thermodynamical variables
$T$ and $\mu$ to be minimal with respect to any free parameters on
which $\Omega$ is dependent. Therefore, the second law of
thermodynamics restricts the class of admissible inequivalent
representations by the condition
\begin{eqnarray}\label{e:secon Law of thermodynamics 4.23}
  \bigg(\frac{\partial\Omega}{\partial(\Delta\Delta^{*})}\bigg)_{T,\mu}
  &=&
  -\frac{\Delta\Delta^{*}}{g}\frac{(1-D)^{2}}{(1-D-2\Delta\Delta^{*}C)^{2}}\nonumber\\
  &&\hspace{-15pt}\times
  \big(
   3C+2\Delta\Delta^{*}\frac{\partial C}{\partial (\Delta\Delta^{*})}
  \big)=0
\end{eqnarray}
 The last condition admits only
two solutions
\begin{subequations}\label{e:two solutions 4.24 and 4.25}
\begin{equation}\label{e:solution 4.24}
  \Delta\Delta^{*}=0
\end{equation}
and $1-D=0,$ i.e.
\begin{equation}\label{e:solution 4.25}
  1=\frac{g}{(2\pi)^{3}}\int\frac{{\rm d}^{3}{k}}{2E_{\bm{k}}}\tanh{\frac{\beta
%
%
  E_{\bm{k}}}{2}}\theta\big(\hbar\omega-|\xi_{\bm{k}}|\big),
%
%
\end{equation}
\end{subequations}
because the expression
$$3C+2\Delta\Delta^{*}\frac{\partial C}{\partial (\Delta\Delta^{*})}$$
is always negative.

The last relation is the well-known gap equation of the BCS theory
of superconductivity \cite{FetterWalecka}. Its solution gives the
gap functions $\Delta$ and $\Delta^{*}$ as certain functions of the
temperature $T$ and the chemical potential $\mu$ at
$T<T_{\text{c}},$ where $T_{\text{c}}$ is the critical temperature.
Thus, the superconducting state of the electron system described by
the action functional (\ref{e:BCS 4.17}) is associated at each value
of $T$ with the corresponding inequivalent representation of the
anticommutator ring of the field operators
$\bm{a}_{\bm{k},\sigma}^{\vphantom{+}}$ and
$\bm{a}_{\bm{k},\sigma}^{+}$ specified by the set of the parameters
$\alpha_{\bm{k}}$ given by (\ref{e:alpha 4.22}).

The result of the functional integral (\ref{e:functional Z}) with
the action functional (\ref{e:BCS 4.17}) as given by (\ref{e:BCS
Omega 4.19}) is not the only one. One can, in fact, also find for
it another class of inequivalent representations of the
anticommutator ring of electron field operators as discussed in
\cite{MatejkaNoga}.

\section{Conclusions}
In quantum field theories, each operator
$A(\bm{a}_{\vphantom{k}}^{+},\bm{a}),$ as a function of the field
operators $\bm{a}_{\bm{k},\sigma}^{+}$ and
$\bm{a}_{\bm{k},\sigma}^{\vphantom{+}}$ satisfying the commutator or
anticommutator ring of the field operators, is an abstract object.
It can be represented in infinitely many inequivalent
representations of the commutator or anticommutator ring of the
field operators. Its trace $\mathop{\rm Tr}
A(\bm{a}_{\vphantom{k}}^{+},\bm{a}),$ is a number which is distinct
for each inequivalent representation.

In the functional integral formalism of quantum field theories, one
associates with each operator $A(\bm{a}_{\vphantom{k}}^{+},\bm{a})$
its kernel $\tilde{A}(a_{\vphantom{k}}^{*}, a).$ The trace of the
operator $A(\bm{a}_{\vphantom{k}}^{+},\bm{a})$ is expressed by the
functional integral \cite{Berezin}
\begin{equation}\label{e:trace 5.1}
  \mathop{\rm Tr}
  A(\bm{a}_{\vphantom{k}}^{+},\bm{a})=\int\tilde{A}(a_{\vphantom{k}}^{*}, a)e^{-a_{\vphantom{k}}^{*} a}\mathscr{D}(a_{\vphantom{k}}^{*}, a),
\end{equation}
which should be regarded as an abstract object. All problems of
quantum field theories based on the functional integral  formalism
can be thought of as problems of finding correct definitions and
computational methods for the functional integrals  of the type
(\ref{e:trace 5.1}). By the selection of a method for the
evaluation of the functional integral (\ref{e:trace 5.1}), one
selects tacitly an inequivalent representation of the commutator
or anticommutator ring of field operators. From this viewpoint,
functional integrals in quantum field theories cannot be regarded
as Newton-Lebesgue integrals giving unique values as one expects
in ordinary integral calculus. Distinct values corresponding to
the same functional integrals in quantum field theories reflect
one of the fundamental properties of such theories, namely, the
existence of infinitely many  inequivalent representations for the
same operator. From this viewpoint, the unexpected properties of
functional integrals in quantum field theories should not be
associated as with their \textit{a priori} undefined expressions,
but with the fundamental structure of quantum field theories.

\section*{Acknowledgements}
The author (M.N.) is very grateful to Prof. C. Cronstr{\"o}m for
many stimulating discussions on the relations between functional
integrals and inequivalent representations in quantum field
theories.

\bibliography{BibJTPh2004TelekiNoga117}
\end{document}